\documentclass{easychair}

\usepackage{doc}
\usepackage{makeidx}

\newcommand{\easychair}{\textsf{easychair}}

\begin{document}

\title{Using of GPUs for cluster analysis of large data by $K$-means method}

\titlerunning{The {\easychair} Class File}

% Authors are joined by \and. Their affiliations are given by \inst, which indexes
% into the list defined using \institute
%
\author{Natalya Litvinenko \inst{}}

% Institutes for affiliations are also joined by \and,
\institute{
Institute of mathematics and mathematical modeling NAS RK\\
Al-Farabi Kazakh National University,  Almaty, Kazakhstan\\
  \email{n.litvinenko@inbox.ru}
 }

\authorrunning{Litvinenko}

\clearpage

%%%%%%%%%%%%%%%%%%%%%%%%%%%%%%%%%%%%%%%%%%%%%%%%%%%
\maketitle
%%%%%%%%%%%%%%%%%%%%%%%%%%%%%%%%%%%%%%%%%%%%%%%%%%%

\begin{abstract}
This problem was solved within the framework of the grant project \textit{Solving of problems of cluster analysis with application of parallel algorithms and cloud technologies} in the Institute of Mathematics and Mathematical Modelling in Almaty.  

The problem of cluster analysis for the large amount of data is very important in different areas of science - genetics, biology, sociology etc. At the same time, such statistical known packages as STATISTICA, STADIA, SYSTAT and others do not allow to solve large problems. The new algorithm that uses the high processing power of GPUs for solving clustering problems by the $K$-means method was developed. This algorithm is implemented as a C++ application in Microsoft Visual Studio 2010 with using the GPU Nvidia GeForce 660. 

The developed software package for solving clustering problems by the method of $K$ - means with using GPUs allows us to handle up to 2 million records with number of features up to 25. The gain in the computing time is in factor 5. We plan to increase this factor up to 20-30 after improving the algorithms.

\end{abstract}

\setcounter{tocdepth}{2}
{\small
\tableofcontents}

\pagestyle{empty}

%------------------------------------------------------------------------------
\section{Introduction}
\label{sect:introduction}

\noindent 
 In this paper we consider the clustering problem with a large number of data ~-~ up to $n=2*10^6$
samples and up to $M=25$ features. The clustering problem is a large multi-step computational problem splitting of a set of objects into groups called clusters. Each cluster consists of similar objects and the objects of different clusters are significantly different. The clusters are constructed by the $K$-means method as follows:\\

\textbf {Algorithm 1}

\begin{enumerate}
\item
Randomly choose $K$ objects which are far away from each other. This selection of objects is arbitrary, but the choice influences on the number of iterations and the computing time. These objects will be the centers of the future clusters $\{S_1,...,S_K\}$ in the first approximation. 
\item
We assign every other object to the cluster whose center is closest. On this stage the first iteration is finished.
\item
Compute the center of gravity for every constructed cluster $S_i$, $i=1..K$,
\begin{equation}
\vec{r}_c : =\frac{\sum_{j=1}^{n_i} \vec{r}_j}{n_i}
\end{equation}
where $n_i$ is the number of objects in the cluster $S_i$ and $\vec{r}_j$ is the radius vector of the $j$-th object.
\item
Assign all the objects to the cluster whose new center is closest to them.
\item
Compute the new centers of gravity of the constructed $K$ clusters.
\item
Compare the centers of gravity of the clusters, formed in the last two iterations.
\item
If all the centers of gravity are congruent, the computations are finished; otherwise we repeat the computations beginning with step 4.
\end{enumerate}

Further in this paper we present three algorithms:
\begin{enumerate}
\item
Single-threaded regime without using GPU.
\item
Multi-threaded regime without using GPU.
\item
Multi-threaded regime with using GPU.
\end{enumerate}

%------------------------------------------------------------------------------
\section{Analysis of recent results}
\label{sect:Analysis of recent results}

There is a large number of statistical packages nowadays. All these statistical packages can be divided into two groups: universal and special-purposed ones. Universal packages offer a wide range of statistical methods for solving various problems without division into subject areas. This group includes such packages as SAS, STATISTICA, MINILAB. In special-purposed packages the number of used statistical methods is restricted by specific subject areas. Sometimes facilities of universal packages are not enough. In that case we should try to apply specialized packages (e.g. STADIA or STATIT). Such specialized packages may require large financial costs. Additionally they may have significant limitations in use, in particular, on the amount of processed data. Other difficulties in using these packages are lack of user-friendly manuals or non-intuitive interface. Because of all there reasons and the large amount of data we decided to develop our own software package for solving the clustering problem. The packages listed above are not able to work with such large data sets.

%------------------------------------------------------------------------------
\section{Similar developments}
\label{sect:Similar developments}

The arising applied problems are often confidential because they are very important in such areas as medicine, genetic engineering and others. Therefore we have only little information that similar software products were developed in the leading Russian research institutions. Thus, we do not have the opportunity to study these products in detail, evaluate their technical characteristics, prospects and results. Unfortunately, similar public domain software products could not be found.

%------------------------------------------------------------------------------

\section{Problem statement}
\label{sect:Problem statement}

The problem of cluster analysis with large data (up to 2 million samples) with the $K$-means method should be solved in three regimes - single-threaded, multi-threaded and multi-threaded with GPUs \cite{gergel}. Please note that to demonstrate the efficiency of the multi-threaded regime (especially with using GPUs) a large amount of data is needed. For a small amount of data, selection of the regime (single-threaded or multi-threaded) should be done automatically. As a first approximation we will assume that a single-threaded regime should be used for problems with less than 10000 samples. In problems with up to 100000 samples, the user should have a choice between a single-threaded and multi-threaded regime. In complexer problems the user should be able to use all three regimes.

%------------------------------------------------------------

\section{General idea of the Algorithm}
\label{sect:General idea of the Algorithm}

The main goal of this work is the research of parallel algorithms with GPUs \cite{sanders} for solving clustering problems. Therefore, we did not consider the problems associated with the correct preparation of the initial data. By default, the distance between $i$-th and $\ell$-th objects is defined as follows:
\begin{equation}
\rho : =\sqrt{ \sum_{j=1}^M (x_{j,i} - x_{j,\ell})^2}, 
\end{equation}
If necessary, other metrics can be chosen. The problem in a single-threaded regime can be solved by the following algorithm:\\

\textbf {Algorithm 2: (problem in single-threaded regime)}

\begin{enumerate}
\item
Compute diameter $D$ of the sample set, 
\begin{equation}
\ D =\mbox{max}_{k,\ell}\left \{\sqrt{ \sum_{j=1}^M (x_{j,k} - x_{j,\ell})^2}\right \}, 
\end{equation} 
i.e. find two objects with the largest distance between them.
\item
Determine the center of gravity $C$ of this data set.
\item
Define $K$ points that will be centers of gravity of clusters in the first approximation.
\item
Assign every object to the cluster whose center is closest. The first iteration is finished.
\item
Find the centers of gravity of the constructed $K$ clusters.
\item
Assign every object to the cluster whose new center is closest.
\item
Find new centers of gravity of the constructed $K$ clusters.
\item
Compare the centers of gravity of the clusters, formed in the last two iterations. If all the centers of gravity are congruent, the computations are finished, otherwise repeat steps 6-8.
\end{enumerate}

Until now all computations were performed in a single-thread regime. The algorithm for solving the clustering problem in a multi-threaded regime is the following:\\

\textbf {Algorithm 3: (in multi-threaded regime)}

Let $N$ be the number of threads.
\begin{enumerate}
\item
Compute diameter $D$ of the sample set.
All computations are performed in the multi-thread regime. In each thread compute distances between the elements of the whole set and elements of ($1/N$)-th part of this set. Find the largest distance in every thread. Then, when all the threads have finished their work, choose a pair of elements with the largest distance between them.
\item
Compute center of gravity $C$ of the whole sample set.  Compute the sum of coordinates of ($1/N$)-th part of the whole set in each thread. When all the threads have finished their work, compute the total sum of coordinates. Then compute the center of gravity.
\item
Define $K$ points that are centers of gravity of clusters.
\item
Assign every object to the cluster whose new center is closest. All computations are performed in the multi-thread regime. Every thread handles ($1/N$)-th part of the elements of the whole set.
\item
Find centers of gravity of the constructed $K$ clusters. All computations are performed in the multi-thread regime.
\item
Assign every object to the cluster whose new center is closest.
\item
Find new centers of gravity of the constructed $K$ clusters.
\item
Compare in the single-threaded regime the centers of gravity of the clusters, formed in the last two iterations.
\item
If all the centers of gravity are congruent, the computations are finished; otherwise repeat steps 6-8.
\end{enumerate}

The solution process in the multi-thread regime with using GPUs is as follow:\\

\textbf {Algorithm 4: (in multi-threaded regime with using GPUs)}
\begin{enumerate}
\item
Compute diameter $D$ of the sample set. All computations are performed in the multi-threads regime. In each thread compute distances between the elements of the whole set and elements of ($1/N$)-th part of this set. Each thread prepares the task for the GPU, sends this task for execution and receives the results – the pair of elements with the largest distance between them. Then, when all the threads have finished their work, choose a pair of elements with the largest distance between them.
\item
Compute center of gravity C of the whole data set. Compute the sum of coordinates of ($1/N$)-th part of the whole set in each thread. Each thread prepares the task for the GPU, sends task for execution and receives the results – the sum of coordinates of elements processed by this thread. When all the threads have finished their work, compute the total sum of coordinates and then the center of gravity.
\item
Define $K$ points that are centers of gravity of clusters.
\item
Assign every object to the cluster whose new center is closest. All computations are performed in the multiple-thread regime. Every thread handles ($1/N$)-th part of the elements of the whole set. GPUs are not used for these tasks. On this stage the first iteration is finished.
\item
Find centers of gravity of the constructed $K$ clusters.
\item
Assign every object to the cluster whose new center is closest.
\item
Find the new center of gravity of the constructed $K$ clusters.
\item
Compare (in the single-threaded regime) the centers of gravity of the clusters, formed in the last two iterations.
\item
If all the centers of gravity are congruent, the computations are finished; otherwise repeat steps 6-8.
\end{enumerate}

\textbf {Intermediate conclusion}: This clustering problem is quite well parallelizable. However, the number of computations is not so large at each stage. The parallelization requires certain computational expenses. The expenses for the CPU parallelization can be covered by the parallel computations. But expenses for the usage of GPUs are not covered by the win of GPU parallelization and sometimes even increase the total computational cost  \cite{boreskov}. The main problem is the insufficient number of computations.

%------------------------------------------------------------------------------
\section{Development environment}
\label{sect:Development environment}

For the software development we used a PC with the motherboard - Gigabyte Technology Co., Ltd., Z77MX-D3H with Intel chipset; CPU - Intel(R) Core(TM) i7-3770 CPU @ 3.40GHz; GPU - NVIDIA GeForce GTX 660 \cite{zib}; RAM 16384 Mb; hard disk - 2 Tb. We used the operating system Microsoft Windows 7, Ultimate, 32 bit, programming environment - Microsoft Visual Studio 2010 on C++ \cite{horton} with using CUDA 5.5.

%------------------------------------------------------------------------------

\section{Future works and plans}
\label{sect:Future works and plans}

The problem of clustering of large data is very important in different areas of science - genetics, sociology, computer science, etc. It may happened that other methods of clustering will be more profitable. Thus, it can be useful to consider other clustering methods – single linkage method, average linkage method, pair-group method using the centroid average, etc. Each of these methods requires an individual approach. The main direction in the development of this topic will be design and implementation of algorithms for new methods as well as implementation of constraints arising in applied tasks. Later on computational efficiency of all, mentioned above, parallel clustering methods will be compared. The next idea is to develop parallel algorithms for the shared memory architecture with the purpose to increase the computational efficiency. It should give a significant gain in computational speed in comparison with the global GPU memory. Additionally, we will also consider the idea of using TESLA GPUs.

%------------------------------------------------------------------------------

\section{Conclusion}
\label{sect:Conclusion}

The construction of clusters by the $K$-means method does not require so many computations as, for example, complete-linkage clustering. Therefore, the usage of GPUs is not always reasonable. In many cases it is enough to use a multi-threaded regime.

%------------------------------------------------------------------------------
% Refs:
%
\label{sect:bib}
\bibliographystyle{plain}
\bibliography{easychair}

%------------------------------------------------------------------------------

\end{document}